\documentclass[a4paper, 10 pt, conference]{ieeeconf}
\IEEEoverridecommandlockouts
\usepackage{comment}
\usepackage{graphics} 
\usepackage{caption}
\usepackage{fancyref}
\usepackage{subcaption}
\usepackage{multirow}
\usepackage{graphicx}
\usepackage{xcolor}
\usepackage{hyperref}
\usepackage{cite}
\DeclareMathAlphabet{\mathcal}{OMS}{cmsy}{m}{n}
\usepackage{epsfig} 
\usepackage{mathptmx} 
\usepackage{times} 
\usepackage{amsmath} 
\usepackage{amssymb}  
\usepackage{pifont}
\newcommand{\zono}[1]{\left\langle#1\right\rangle}

\newtheorem{definition}{Definition}
\newtheorem{proposition}{Proposition}

\newcommand{\R}{{\mathbb{R}}}

\title{\LARGE \bf Set-Membership Estimation in Shared Situational Awareness\\ for Automated Vehicles in Occluded Scenarios}

\author{Vandana Narri$^{1,2}$, Amr Alanwar$^{2}$, Jonas Mårtensson$^{2}$, Christoffer Norén$^{1}$,\\
Laura Dal Col$^{1}$ and Karl Henrik Johansson$^{2}   $
\thanks{This paper is submitted to IEEE Intelligent Vehicles Symposium 2021.}
\thanks{$^{1}$The authors are with Research and Development, Scania CV AB, 151 87 Södertälje, Sweden. {\tt\small \{vandana.narri, christoffer.noren, laura.dal.col\}@scania.com.}}%
\thanks{$^{2}$The authors are with Division of Decision and Control Systems at School of Electrical Engineering and Computer Science, KTH Royal Institute of Technology, SE100 44 Stockholm, Sweden. {\tt\small\{narri, alanwar, jonas1, kallej\}@kth.se.}}%
}

\begin{document}

\maketitle

\begin{abstract}
One of the main challenges in developing autonomous transport systems based on connected and automated vehicles is the comprehension and understanding of the environment around each vehicle. In many situations, the understanding is limited to the information gathered by the sensors mounted on the ego-vehicle, and it might be severely affected by occlusion caused by other vehicles or fixed obstacles along the road. Situational awareness is the ability to perceive and comprehend a traffic situation and to predict the intent of vehicles and road users in the surrounding of the ego-vehicle. The main objective of this paper is to propose a framework for how to automatically increase the situational awareness for an automatic bus in a realistic scenario when a pedestrian behind a parked truck might decide to walk across the road. Depending on the ego-vehicle's ability to fuse information from sensors in other vehicles or in the infrastructure, shared situational awareness is developed using a set-based estimation technique that provides robust guarantees for the location of the pedestrian. A two-level information fusion architecture is adopted, where sensor measurements are fused locally, and then the corresponding estimates are shared between vehicles and units in the infrastructure. 
Thanks to the provided safety guarantees, it is possible to appropriately adjust the ego-vehicle speed to maintain a proper safety margin. It is also argued that the framework is suitable for handling sensor failures and false detections in a systematic way. Three scenarios of growing information complexity are considered throughout the study. Simulations show how the increased situational awareness allows the ego-vehicle to maintain a reasonable speed without sacrificing safety. 
\end{abstract}

\section{Introduction}
\begin{figure}[t]
    \centering
    \includegraphics[width=0.9\linewidth]{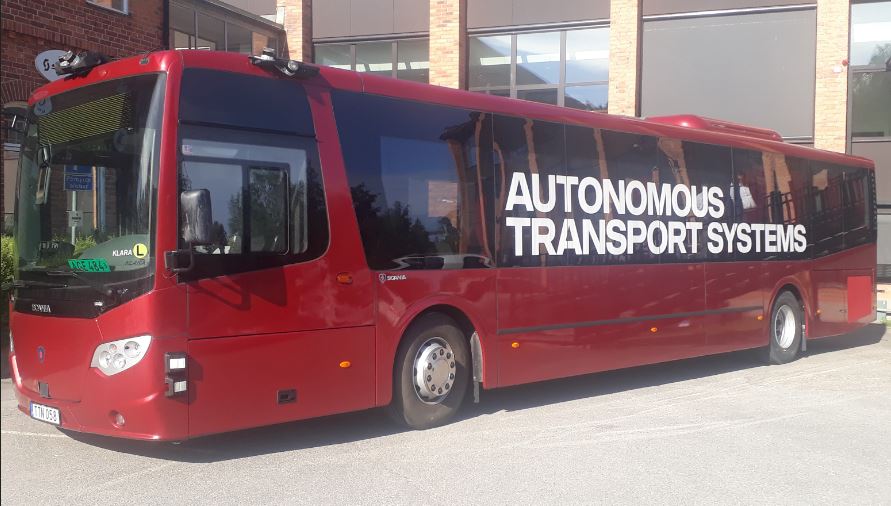}
    \caption{Autonomous Scania bus equipped with lidar, radar, camera and other sensors.}
    \label{fig:autonomous_bus}
\end{figure}

Connected and automated vehicles (CAVs) have attracted enormous attention in the last decade because of their expected impact on the economy and society in general. The benefits of such vehicles range from transforming the travel experience, improving safety, and reducing environmental impact to enabling more competitive road freight transportation and novel public transport modalities~\cite{paden_survey_2016,gonzalez_review_2016}. When automating the driving of CAVs, the ability of the vehicles to assess and reason about their state and the nearby environment is of utmost importance. A vehicle's situational awareness becomes crucial for handling challenging scenarios, such as occlusion due to other vehicles parked on the side of a road or moving in front of the ego-vehicle. In this paper, we consider a particular set of such scenarios that are likely to arise for an automated public bus in an urban setting, such as the autonomous Scania bus, equipped with lidar, radar, camera, and other sensors, illustrated in  Fig.~\ref{fig:autonomous_bus} and to be used in the future to implement and evaluate the approach developed in this paper.   

There is a dramatic increase of interest in situational awareness for CAVs. Questions often considered include how to estimate the states of pedestrians or bicyclists present in the surrounding of the ego-vehicle. When there are multiple CAVs present, shared situational awareness is a relevant concept~\cite{grembek2019making, barnett2020automated, gopalswamy2018infrastructure,moradi2017utilizing, lenz2016tactical}. Statistical approaches are commonly considered, e.g., the authors of~\cite{simon2010kalman} use a Kalman filter to estimate the state of the actor and assumes linear models and Gaussian noise. A relevant review of constrained Bayesian state estimation for linear and nonlinear state-space systems is given in~\cite{amor2018constrained}. Most existing methods for situational awareness, consider traditional point-wise state estimators~\cite{thrun_probabilistic_2006}. 


Safety guarantees are often well characterized by robust or set-based estimators~\cite{rego2018set}. One of the most popular set-based approaches is set-membership estimators. Such estimators have intuitive geometric processing in that they intersect the set of states consistent with the model and the set consistent with the measurements to obtain the corrected state set~\cite{conf:dis-diff}. The estimated set can be mathematically characterized as ellipsoid~\cite{bolting2019iterated,merhy2020guaranteed,bolting2018set, jaulin2009nonlinear} or zonotopes~\cite{garcia2020guaranteed,conf:dis-diff}. The suitable choice is application dependent. 
Set-membership estimators are used in many applications such as fault detection~\cite{conf:faultcombastel1,conf:set_fault1}, underwater robotics~\cite{conf:setmem2water, jaulin2009nonlinear}, ground vehicles~\cite{franze2015obstacle}, multi-agent systems~\cite{conf:setmem3eventleader,bolting2019iterated,garcia2020guaranteed,bolting2018set}, and localization~\cite{conf:setloc}. Self-localization has been considered by many researchers, for autonomous robots~\cite{di2003simultaneous}, drones~\cite{merhy2020guaranteed}, and other vehicles~\cite{bento2018set}. It has been shown that the method is particularly advantageous in a dynamic unknown environment as~\cite{di2001set,garulli2001set,rohou2017guaranteed}. 



The main contribution of this paper is an approach for shared situational awareness aimed at improving the perception and operation of an automated heavy-duty vehicle by letting it systematically process information from multiple dynamic sensors in the environment. A setup is considered with an automatic bus driving along a road when a pedestrian behind a parked truck might decide to walk across the road. The ego-vehicle (the bus) is able to fuse information from one or more sensors located on other vehicles or infrastructure units. A novel situational awareness framework is developed using set-based estimation techniques that provide robust guarantees for the pedestrian's location. A two-level information fusion is adopted, where sensor measurements are fused locally, and then the corresponding estimates are shared between vehicles and the infrastructure units. 

Zonotopes are used to represent the sets in which the true states are guaranteed to belong. It is shown that this representation allows an efficient computation for sensor fusion and state estimation and that information from one or more units can be easily incorporated into the estimated set. Thanks to the provided safety guarantees, it is possible to appropriately adjust the speed of the ego-vehicle to always maintain a proper safety margin. Three scenarios of growing information complexity are considered, where in the first scenario the ego-vehicle uses only its own sensors, in the second scenario it incorporates information also from another vehicle using V2V communication, and in the third scenario two sensors at the road crossings are utilized. Simulations show how the increased situational awareness allows the ego-vehicle to maintain a reasonable speed without sacrificing safety. 

The remaining part of the paper is outlined as follows. In Section~\ref{sec:pb}, the problem formulation is described together with the three scenarios considered throughout the paper. Section~\ref{sec:mehto} presents the situational awareness framework based on set-based estimation. An extensive simulation study is performed in Section~\ref{sec:eval} illustrating the approach and its limitations. Finally, concluding remarks are given in Section~\ref{sec:conc} together with a brief discussion on future directions.

\section{Problem Formulation}
\label{sec:pb}

\begin{figure}[t]
\centering
   \begin{subfigure}[b]{0.5\textwidth}
   \includegraphics[trim={0 0 0 0},clip,width=0.9\linewidth,height=0.55\linewidth]{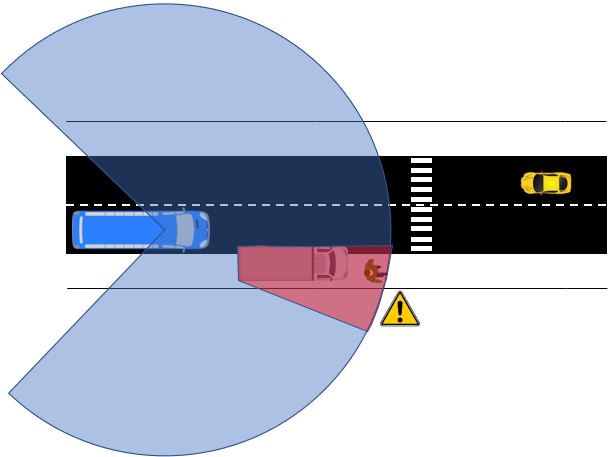}
    \caption{Scenario 1: The ego-vehicle with one local sensor is approaching the zebra crossing. The field of view is obstructed by a parked truck. A pedestrian is in the occluded region.}
    \label{fig:urban_scenario} 
\end{subfigure} 
\begin{subfigure}[b]{0.5\textwidth}
   \includegraphics[trim={0 1cm 0 1cm},clip,width=1\linewidth,height=0.5\linewidth]{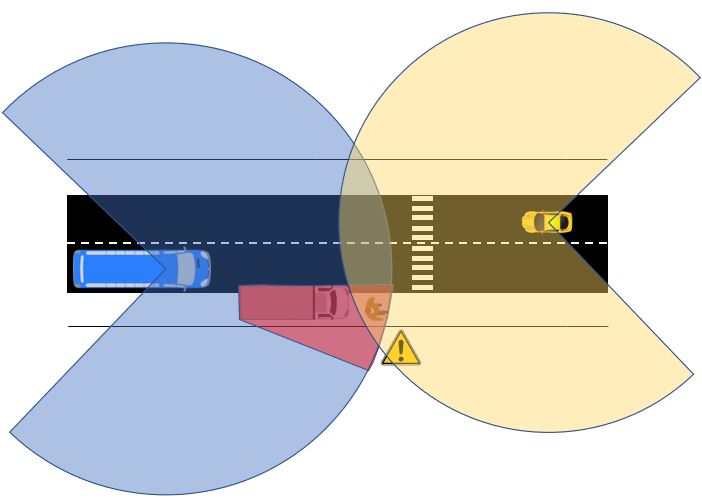}
    \caption{Scenario 2: While the ego-vehicle is approaching the zebra crossing, it connects with an approaching vehicle equipped with another sensor. The pedestrian is visible by the approaching vehicle's sensor.}
    \label{fig:urban_scenario_with_cv}
\end{subfigure}
\begin{subfigure}[b]{0.5\textwidth}
   \includegraphics[trim={0 2cm 0 1cm},clip,width=0.85\linewidth,height=0.4\linewidth]{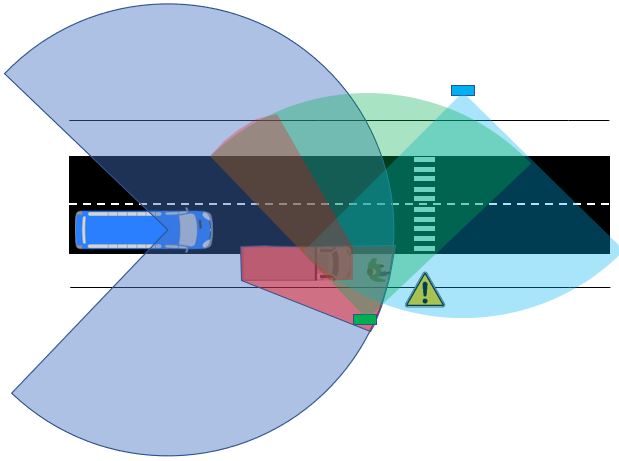}
    \caption{Scenario 3: The ego-vehicle connects to two road-side units with one sensor each. These sensors cover the zebra crossing by their field of view.}
    \label{fig:urban_scenario_with_roadside}
\end{subfigure}
\caption{The three considered scenarios of an connected and automated bus passing a zebra crossing.}
\label{fig:scenarios} 
\end{figure}

The problem considered in this paper is formulated around the three scenarios in Fig.~\ref{fig:scenarios}. These scenarios consist of a two-lane road with a side-walk on each side of the road and a zebra crossing. The ego-vehicle (blue bus) is traveling from left to right and is approaching the crossing. The ego-vehicle is equipped with a sensor having a field of view represented by the blue-shaded circle segment. As indicated, the field of view is occluded by a red truck parked at the side of the road. The red shaded region represents the ego-vehicle’s occluded sensor view. 

Scenario~1 in Fig.~\ref{fig:scenarios} corresponds to the nominal case when the information processed in the ego-vehicle is limited to its own sensors. Due to the occlusion, the ego-vehicle is not able to detect potential pedestrians on the side-walk behind the truck. Consequently, the ego-vehicle has to lower its speed to be cautious of an unforeseen actor in its near surrounding. From the perspective of the ego-vehicle, this uncertainty leads to an increased travel time and thereby lower efficiency. 

Scenario~2 in Fig.~\ref{fig:scenarios} is improving this situation by connecting to an approaching CAV equipped with another sensor. The measurements from this sensor can be used for sensor fusion by the ego-vehicle and thereby allows it to conclude if there is a pedestrian in the occluded area or not. A limitation of this approach is obviously that there might not be a connected vehicle exactly when needed. 

Scenario~3 in Fig.~\ref{fig:scenarios} handles this case by incorporating connected road-side sensor units. The sensors cover both sides of the zebra crossing and will thereby detect any pedestrians in the scene. 

With this scenarios in mind, we can formulate the problem solved in this paper as the following questions:
\begin{enumerate}
    \item How to obtain and fuse data from local and external sensors to improve the situational awareness for the ego-vehicle to take appropriate actions?
    \item How to robustly share the estimates to guarantee safety under varying uncertainties in the measurements?
    \item How to quantitatively measure the improvement in the state estimates using set-membership techniques?
\end{enumerate}

\section{Methodology}\label{sec:mehto}

In this section, the architecture of the proposed framework is discussed, and each module of the framework is explained in detail. The proposed framework extends the conventional control architecture of an automated vehicle~\cite{gonzalez_review_2016} by using the set-membership method instead of statistical estimation for sensor fusion. 

\subsection{Architecture}
\begin{figure}[h!]
    \centering
    \includegraphics[width=0.8\linewidth]{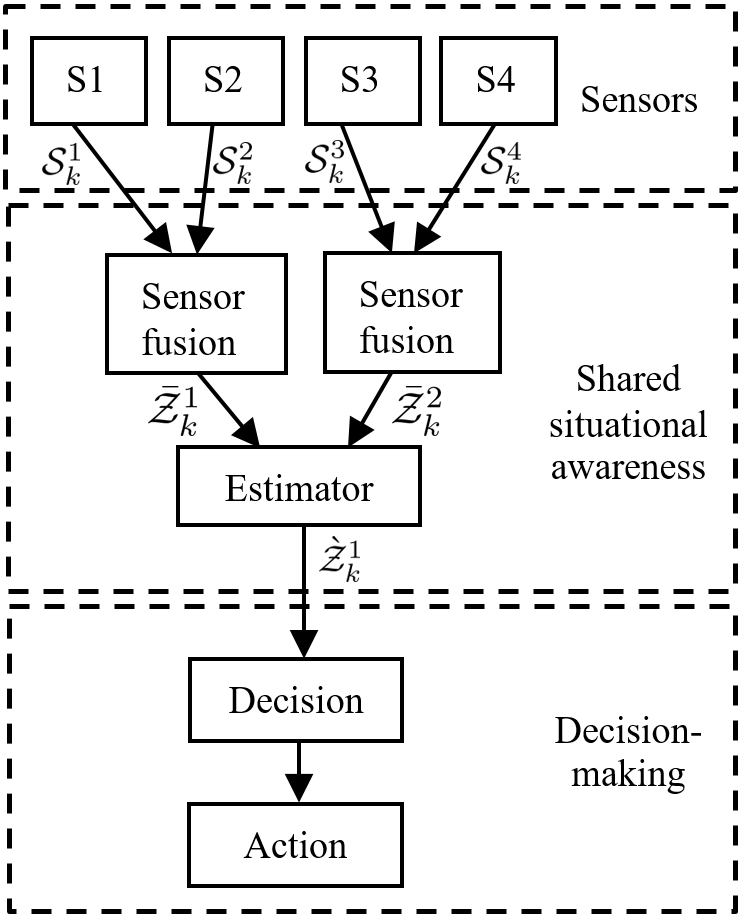}
    \caption{Proposed architecture for set-based estimation for shared situational awareness.}
    \label{fig:flow_chart}
\end{figure}

The proposed architecture is presented in Fig.~\ref{fig:flow_chart}. It consists of three parts: (i) Local and extended sensor network, (ii) Algorithms for shared situational awareness, and (iii) Decision-making. 
Measurement data from the sensors are collected and fused to perform state estimation. Based on these estimates, decisions are made, and actions are planned. In this paper, the main focus is on (i) and (ii).

\subsection{Sensor and System Models}

Each sensor measurement is generated with respect to the orientation of the sensor. The model for sensor $i=1,\dots,4$ is
\begin{equation}
    y^{i}_k = H^i x_k + v_k^i \in \mathbb{R}^{p}
    \label{eq:senmodel}
\end{equation}
for $k = 1,2,\dots, N$, where $N$ is the horizon, $H^i$ the measurement matrix and $v_k^i$ the measurement noise. We assume that the sensor measurements are generated by an observable discrete-time linear system:
\begin{equation}
 x_{k+1} = F x_k + n_k,
\label{eq:sysmodel}
\end{equation}
where $x_k \in \mathbb{R}^{n}$ is the state of the actor, $F$ the state matrix, and $n_k$ process noise.

\subsection{Algorithm for Shared Situational Awareness}

Set-membership estimation with zonotopes is used to estimate $x_k$. Properties of zonotopes are stated next.

\begin{definition}[\textbf{Zonotope \cite{conf:zono1998}}] A zonotope $\mathcal{Z}= \zono{ c,G }$ consists of a center $c \in \R^n$ and a generator matrix $G$ $\in$ $\R^{n \times e}$. We compose $G$ of $e$ generators $g^{(i)} \in \R^n$, $i=1,..,e$, where $G=[g^{(1)},...,g^{(e)}]$ and $\beta_i$ is a scaling factor. Hence, 

\begin{equation}
     \mathcal{Z} = \Big\{ c + \sum\limits_{i=1}^{e} \beta_i  g^{(i)} \Big| -1\leq \beta_i \leq 1 \Big\}.
    \label{equ:zonoDef}
\end{equation}
\end{definition}

\begin{figure}[h!]
\begin{subfigure}[b]{0.15\textwidth}
  \centering
  \includegraphics[width=0.5\linewidth]{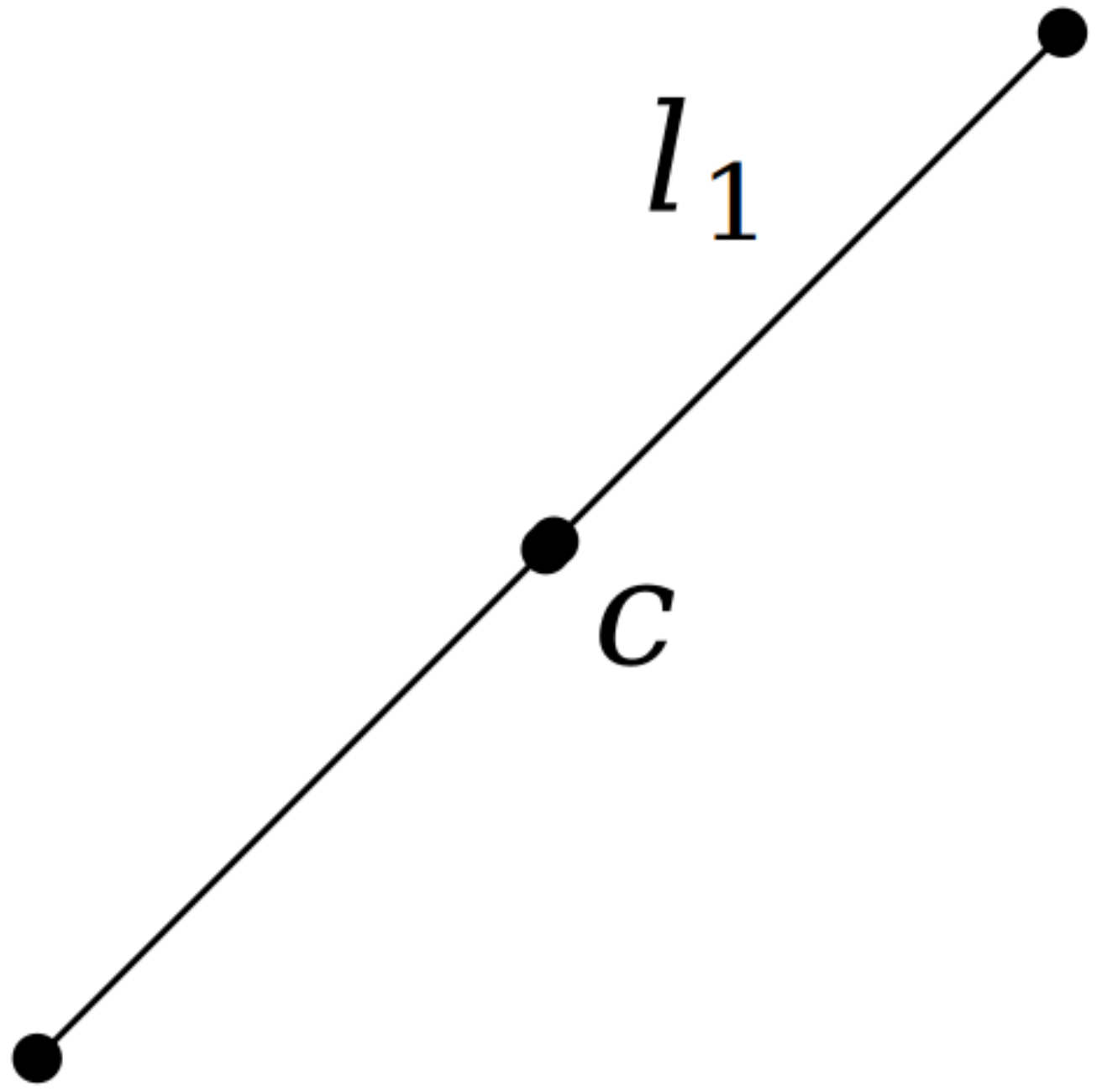}
  \caption{$c \oplus l_1$}
  \label{fig:zono1}
\end{subfigure}%
\begin{subfigure}[b]{0.15\textwidth}
  \centering
  \includegraphics[width=0.5\linewidth]{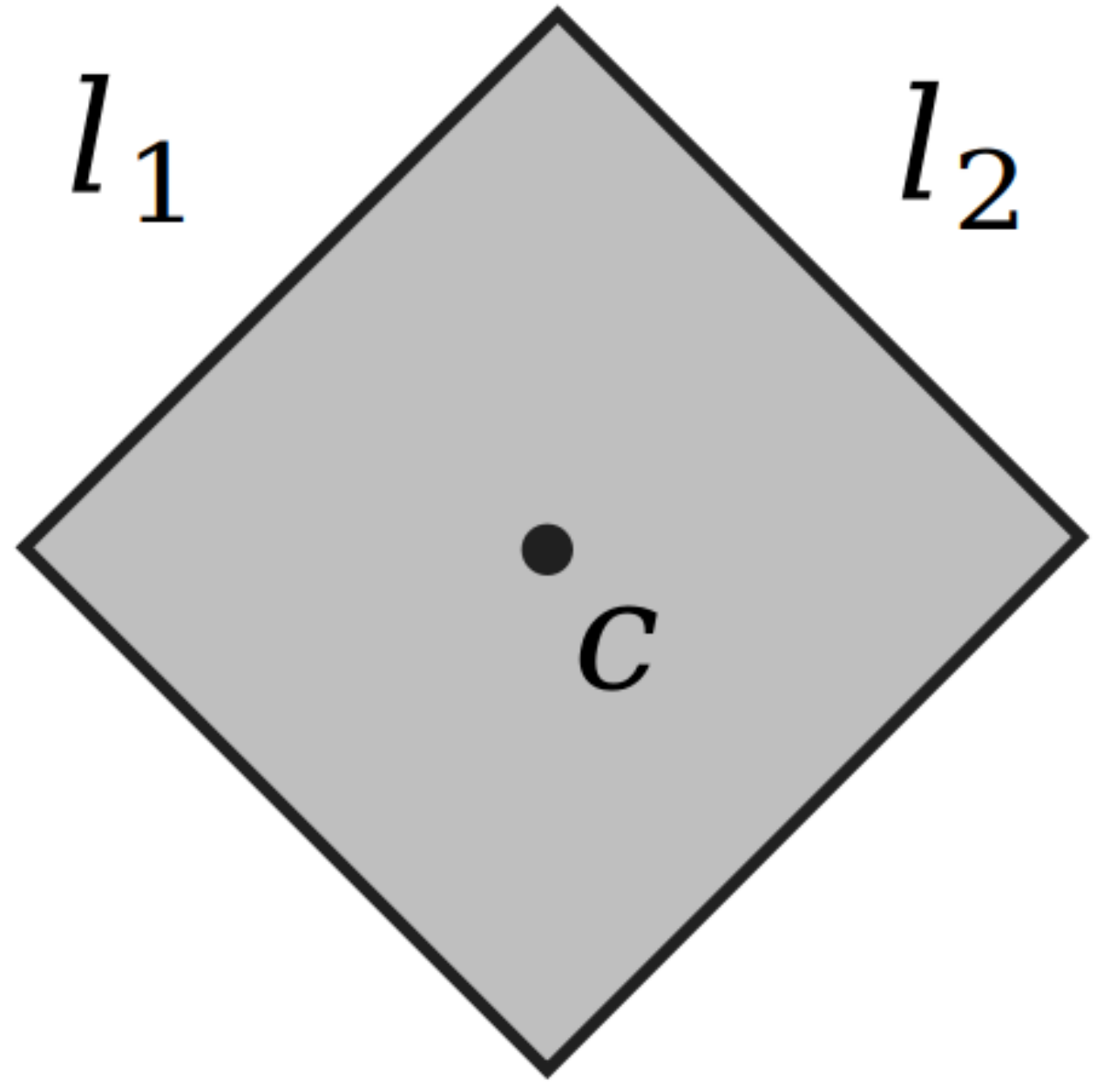}
  \caption{$c \oplus l_1 \oplus l_2$ }
  \label{fig:zono2}
\end{subfigure}
\begin{subfigure}[b]{0.15\textwidth}
  \centering
  \includegraphics[width=0.5\linewidth]{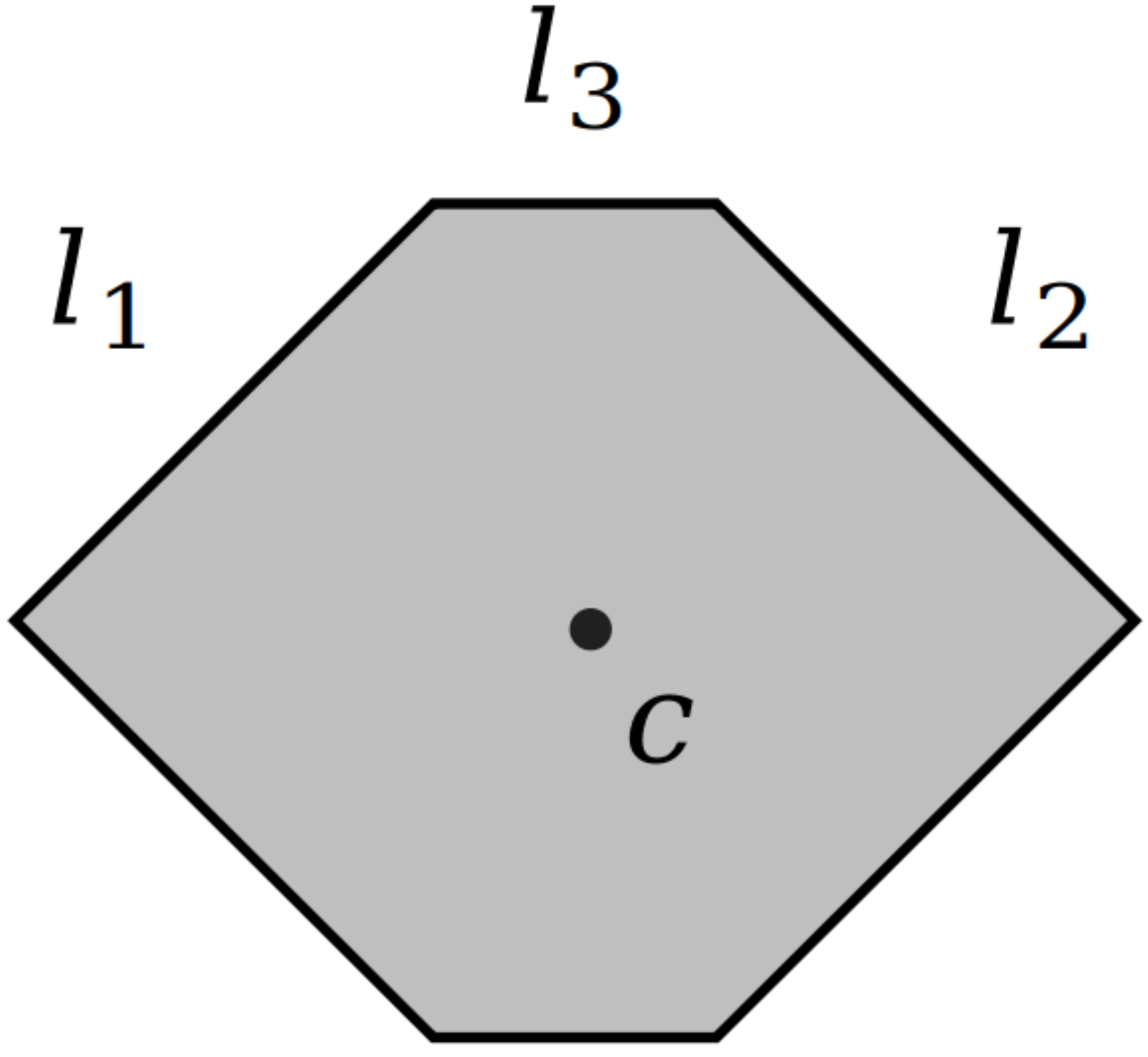}
  \caption{$c \oplus l_1 \oplus l_2 \oplus l_3$}
  \label{fig:zono3}
\end{subfigure}%
\caption{Construction of a zonotope.}
\label{fig:zonocont}
\vspace{-3mm}
\end{figure}

Fig.~\ref{fig:zonocont} shows the construction of a zonotope. Given two zonotopes $\mathcal{Z}_1=\langle c_1,G_1 \rangle$ and $\mathcal{Z}_2=\langle c_2,G_2 \rangle$ and a scalar $L$, the following operations can be computed exactly \cite{conf:zono1998}:
\begin{itemize}
    \item Minkowski sum:
    \begin{equation}
     \mathcal{Z}_1 \oplus \mathcal{Z}_2 = \Big\langle c_1 + c_2, [G_1, G_2]\Big\rangle.
     \label{eq:minkowski}
     \end{equation}
    
    \item Scaling:
    \begin{equation}
     L \mathcal{Z}_1  = \Big\langle L c_1, L G_1\Big\rangle. 
     \label{eq:linmap}
     \end{equation}    
\end{itemize}

\begin{figure}[t!]
    \centering
    \includegraphics[width=0.8\linewidth]{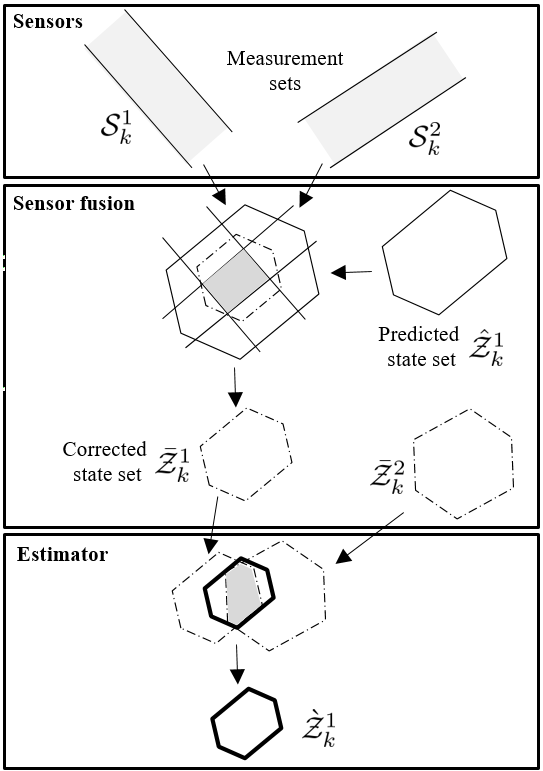}
    \caption{Sensor fusion and estimator.}
    \label{fig:zonostripinter}
\end{figure}

\subsubsection{Sensor Fusion}

Set-membership approach estimation compute set of states instead of a single state. The prediction set is estimated using the model expressed in~\eqref{eq:sysmodel}. The process and measurement noise are assumed to be unknown but bounded by zonotopes: $n_k \in \mathcal{Z}_{Q,k}$, and $v_k^i \in \mathcal{Z}_{R,k}^i$. Then, they intersect the predicted set with the set that aligns with the measurement set. We have the following three sets.

\begin{definition}[\textbf{Predicted State Set}]
\label{def:predset}
Given system~\eqref{eq:senmodel} -- \eqref{eq:sysmodel} with initial set $\mathcal{Z}_{0}= \langle c_{0},G_{0} \rangle$, the predicted reachable set of states $\hat{\mathcal{Z}}_{k}^i$ with noise zonotope  $\mathcal{Z}_{Q,k}$ is:
\begin{equation}
\hat{\mathcal{Z}}_{k}^i= F \hat{\mathcal{Z}}_{k-1}^i \oplus \mathcal{Z}_{Q,k}. 
\end{equation}
\end{definition}

\begin{definition}[\textbf{Measurement State Set}] 
\label{def:measset}
Given system~\eqref{eq:senmodel} -- \eqref{eq:sysmodel}, the measurement state set $\mathcal{S}^i_{k}$ of node $i$ is the set of all possible solutions $x_{k}$ which can be reached given $y_k^i$ and $v_k^i \in \mathcal{Z}_{R,k}^i= \langle0,R_k^i\rangle$ where $R^i_k = \text{diag}([r^1_k, \dots,r^{m_s}_k])$. Note that $y_k^i \in \mathbb{R}^{p}$ is scalar, i.e., $p=1$, this measurement set is a strip:
\begin{equation}
    \mathcal{S}^i_k = \Big\{ x_k \Big| | H^i x_k - y^i_k| \leq r^i_k \Big\}. \label{eq:strip}
\end{equation}
\end{definition}
\begin{definition}[\textbf{Corrected State Set}]
\label{def:corrset}
Given system~\eqref{eq:senmodel} -- \eqref{eq:sysmodel} with initial set $\mathcal{Z}_{0}= \langle c_{0},G_{0} \rangle$, the reachable corrected state set $\bar{\mathcal{Z}}_{k}^i$ of node $i$ is defined as an over approximation of the intersection between $\hat{\mathcal{Z}}_{k}^i$ and $\mathcal{S}^i_k$:
\begin{equation}
     \big( \hat{\mathcal{Z}}_{k}^i \cap \mathcal{S}^i_k \big) \subseteq \bar{\mathcal{Z}}_{k}^i. 
\end{equation}
\end{definition}

Set-membership approaches intersect the set of states consistent with the model  (predicted state set), denoted by $\hat{\mathcal{Z}}_{k-1}^i$, and the sets consistent with the measurements  (measurement state set), denoted by $\mathcal{S}^i_k$, $i = 1,\dots,m_s$, to obtain the corrected state set, denoted by $\bar{\mathcal{Z}}^i_{k}$, also referred as zonotopic set. Strips represent measurements from the sensors and are fused together to get a corrected state set. To fuse the measurements from the sensors and extended sensors, we use of the following proposition.

\begin{proposition}[\cite{conf:stripzono}]
 \label{th:stripzono}
 Given are zonotope $\hat{\mathcal{Z}}_{k-1}^i= \langle \hat{c}^i_{k-1},$ $\hat{G}^i_{k-1} \rangle $, the family of $m_s$ measurement sets $\mathcal{S}^i_k$ in \eqref{eq:strip} and the design parameters $\lambda_{k}^{i,j} \in \R^{n \times p}$, $j = 1,\dots,m_s$. The intersection between the zonotope and measurement sets can be over-approximated by the zonotope $\bar{\mathcal{Z}}^i_{k}=\langle  \bar{c}^i_{k},\bar{G}^i_{k}\rangle $, where
 \begin{align}
  \bar{c}^i_{k} &=  \hat{c}^i_{k-1} + \sum\limits_{j=1}^{m_s} \lambda_{k}^{i,j}(y^{j}_k - H^j \hat{c}^i_{k-1}), \label{eq:C_lambda}\\
  \bar{G}^i_{k} &= \Big[ (I - \sum\limits_{j}^{m_s} \lambda_{k}^{i,j} H^{j} ) \hat{G}^i_{k-1}, \lambda_{k}^{i,1} r^1_k,\dots,\lambda_{k}^{i,m_s} r^{m_s}_k \Big]. 
  \label{eq:G_lambda}
 \end{align}
  \end{proposition}

The design parameter $\lambda_{k}^{i,j}$ can be obtained by solving an optimization problem to minimize the size of the resultant zonotope~\cite{conf:dis-diff }. After fusing the measurements from the sensors, we diffuse the estimates from multiple vehicles and infrastructure units.

\subsubsection{Estimator}

Consider when a road-side unit or connected vehicle are reporting estimates of the same pedestrian. Then, we make use of the following proposition to fuse the set estimates $\bar{\mathcal{Z}}^i_{k}$, $i = 1,\dots,m_e$, by finding their intersections.  
 \begin{proposition}[\cite{conf:dis-diff}]
 \label{th:diff}
The intersection between $m_e$ zonotopes $\bar{\mathcal{Z}}^j_{k}=\zono{\bar{c}^j_{k} , \bar{G}^j_{k}}$ can be over-approximated using the zonotope $\grave{\mathcal{Z}}^i_{k}=\zono{\grave{c}^i_{k},\grave{G}_{k}^i}$ as follows:
 \begin{eqnarray}
\grave{c}^i_{k}&=&\frac{1}{\sum\limits_{j}^{m_e}w^{i,j}_k}\sum\limits_{j}^{m_e}w^{i,j}_k\bar{c}_{k}^j,\label{eq:cdiff}\\
\grave{G}_{k}^i &=& \frac{1}{\sum\limits_{j=1}^{m_e}w^{i,j}_k}[w^{i,1}_k\bar{G}^1_{k} ,...,w^{i,m_e}_k\bar{G}^{m_e}_{k}], \label{eq:gdiff}
 \end{eqnarray}
 where $w^{i,j}_k$ is a weight such that $\sum\limits_{j}^{m_e} w^{i,j}_k \neq 0$.
 \end{proposition}

Again, the design parameter $w^{i,j}_k$ can be obtained by solving an optimization problem to minimize the size of the resultant zonotope \cite{conf:dis-diff}. 

\setlength\tabcolsep{10pt}
\begin{table*}[h!]
    \centering
    \normalsize
    \caption{Comparison of average speed of the ego-vehicle in different scenarios with and without situational awareness.}
    \begin{tabular}{ c c c c c c c c }
    \hline
     Scenario &  Shared situational  &Pedestrian &Local  & Connected  & Road-side  & Road-side  & Average   \\
           &  awareness & & sensor &  vehicle &  unit 1 & unit 2 &  speed \\
        \hline
        \multirow{2}{*}{1} & \multirow{2}{*}{-} & \checkmark & \checkmark & - & - & - & 13.0 \\ 
         &  & - & \checkmark & -  & - & - & 15.0 \\ 
        \hline
        \multirow{2}{*}{2}& \multirow{2}{*}{\checkmark} & \checkmark &\checkmark & \checkmark & - & - & 13.1\\  
         & & - & \checkmark & \checkmark & - & - & 20.0 \\
        \hline
        \multirow{2}{*}{3} & \multirow{2}{*}{\checkmark} & \checkmark & \checkmark & - & \checkmark & \checkmark & 13.2 \\ 
        &  & - & \checkmark & - & \checkmark & \checkmark & 20.0 \\
        \hline
    \end{tabular}
    \label{tab:averagespeed_comparison}
\end{table*}

\subsection{Robust Decision Making}

\begin{figure}[t]
    \centering
    \includegraphics[scale=0.2]{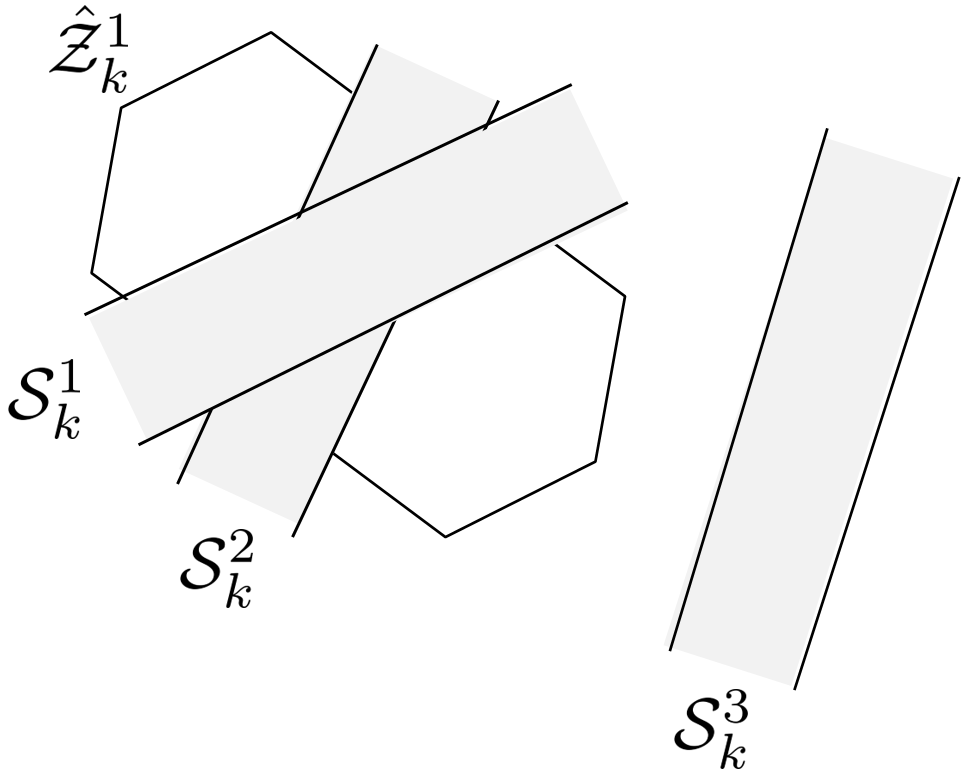}
    \caption{Sensor Failure case after reporting multiple of $\mathcal{S}_k^3$ that does not intersect with the predicted set $\hat{\mathcal{Z}}_k^1$ and other reported measurements ($\mathcal{S}_k^1$ and $\mathcal{S}_k^2$).}
    \label{fig:failure}
\end{figure}

The output of the estimator module acts as the input to the decision-making module. Due to the set-membership approach, a robust decision making can be done. Increasing the number of sensor measurements improves the estimation accuracy. This framework can also detect if there any sensor failures as shown in Fig.~\ref{fig:failure}. 
In this paper, the decision is simply to lower the velocity depending on the uncertainty of the situation. 

\section{Results}\label{sec:eval}

In this section, results of different case studies using the shared situational awareness framework are presented and compared. The ego-vehicle is equipped with four levels of speed, which are stated as (i) Nominal speed - executes 100~\% of the given speed, (ii) Cautious speed  - executes 30~\% of the nominal speed, (iii) Slow speed - executes 50~\% of the nominal speed and (iv) Very slow speed - executes 80~\% of the nominal speed. The different scenarios considered are as stated below: 
\begin{enumerate}
    \item Ego-vehicle with one local sensor with no shared situational awareness.
    \item Ego-vehicle with one local sensor and a connected vehicle with one extended sensor with shared situational awareness.
    \item Ego-vehicle with one local sensor and two road-side units with shared situational awareness.
\end{enumerate}

If there is an occlusion detected in the local sensor's field of view, then the speed of the ego-vehicle is changed from nominal to cautious speed. When a zebra crossing is detected, the speed is decreased to slow speed. The same action is taken if the occlusion is covering the side-walk. Moreover, if a pedestrian or actor is detected in close proximity to the ego-vehicle, it shifts the speed to a very slow level to avoid any collision or accident. Furthermore, when the pedestrian or actor is too close to the ego-vehicle and the decision-making horizon too short, then the steering wheel and brakes of the ego-vehicle are controlled through a reactive or proactive manner by applying emergency brakes.

\begin{figure}
    \centering
    \begin{subfigure}[b]{0.5\textwidth}
        \centering
          \begin{subfigure}[b]{0.45\textwidth}
            \renewcommand\thesubfigure{\alph{subfigure}1}
            \centering
            \includegraphics[trim={2cm 2cm 0 1cm},clip,width=1\linewidth,height=0.7\linewidth]{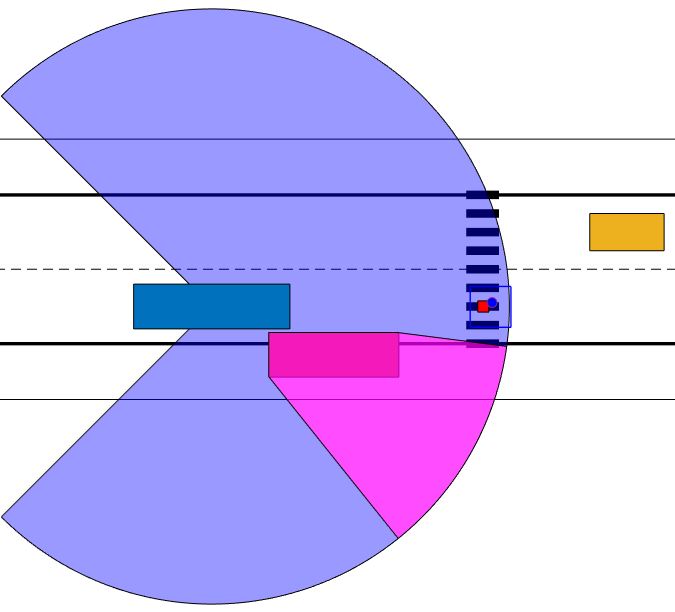}
            \caption{Scenario 1 with pedestrian.}
            \label{fig:scenario1a} 
          \end{subfigure}
          \begin{subfigure}[b]{0.45\textwidth}
              \addtocounter{subfigure}{-1}
              \renewcommand\thesubfigure{\alph{subfigure}2}
              \centering
            \includegraphics[trim={0 1cm 5cm 1cm},clip,width=1\linewidth,height=0.7\linewidth]{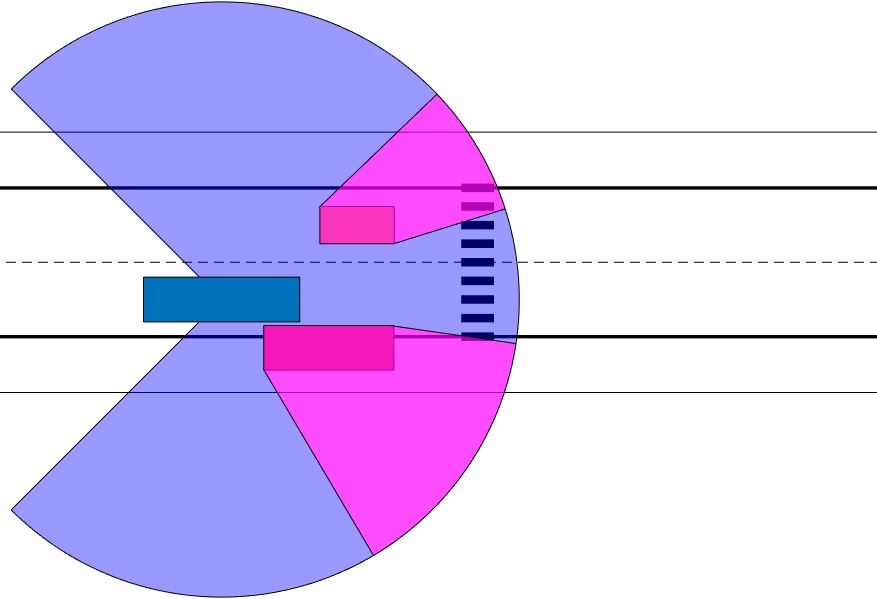}
            \caption{Scenario 1 without pedestrian.}
            \label{fig:scenario1b}
          \end{subfigure}
          \addtocounter{subfigure}{-1}
          \caption{Scenario 1 simulation results}
          \label{fig:scenario1}
    \end{subfigure}
    \begin{subfigure}[b]{0.5\textwidth}
    \centering
     \begin{subfigure}[b]{0.45\textwidth}
        \renewcommand\thesubfigure{\alph{subfigure}1}
        \centering
        \includegraphics[trim={1cm 1cm 2cm 1cm},clip,width=1\linewidth,height=0.7\linewidth]{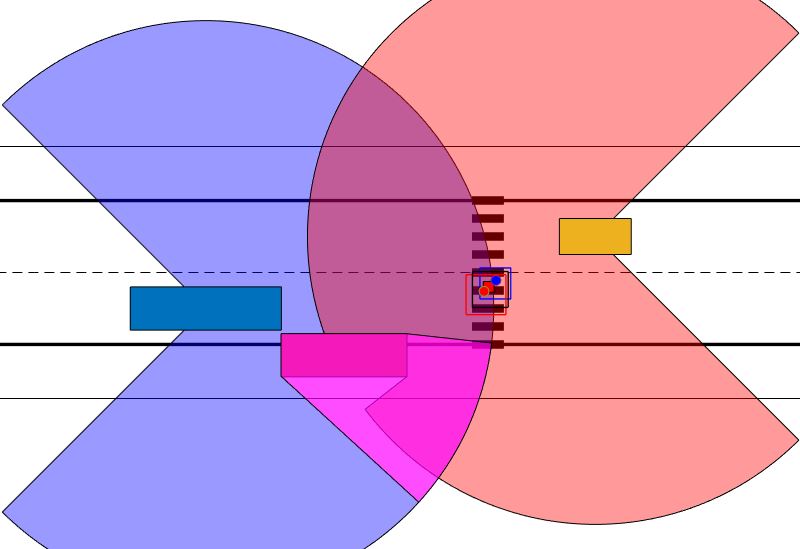}
        \caption{Scenario 2 with pedestrian.}
        \label{fig:scenario2a}
     \end{subfigure}
     \begin{subfigure}[b]{0.45\textwidth}
        \addtocounter{subfigure}{-1}
        \renewcommand\thesubfigure{\alph{subfigure}2}
        \centering
        \includegraphics[trim={1.5cm 1cm 1cm 1cm},clip,width=1\linewidth,height=0.7\linewidth]{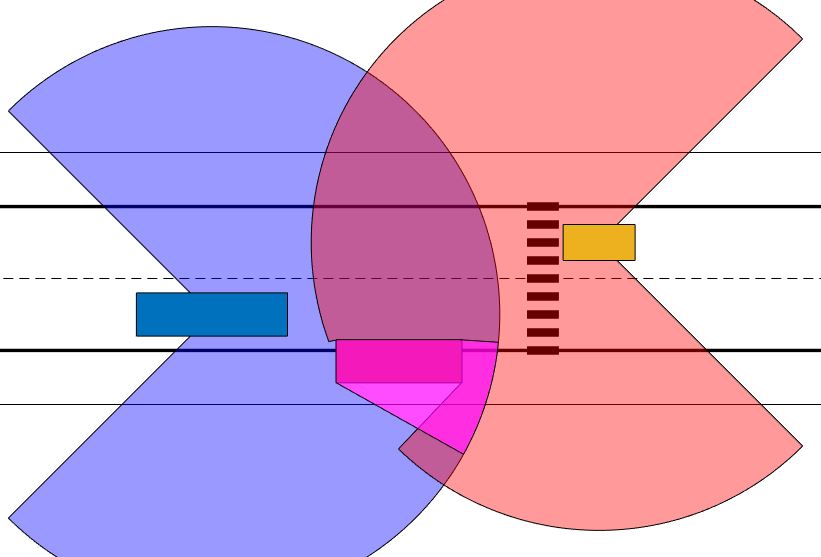}
        \caption{Scenario 2 without pedestrian.}
        \label{fig:scenario2b}
     \end{subfigure}
     \addtocounter{subfigure}{-1}
     \caption{Scenario 2 simulation results}
     \label{fig:scenario2}
    \end{subfigure}
    \begin{subfigure}[b]{0.5\textwidth}
    \centering
      \begin{subfigure}[b]{0.45\textwidth}
       \renewcommand\thesubfigure{\alph{subfigure}1}
        \centering
        \includegraphics[trim={1cm 1cm 3cm 1cm},clip,width=1\linewidth,height=0.7\linewidth]{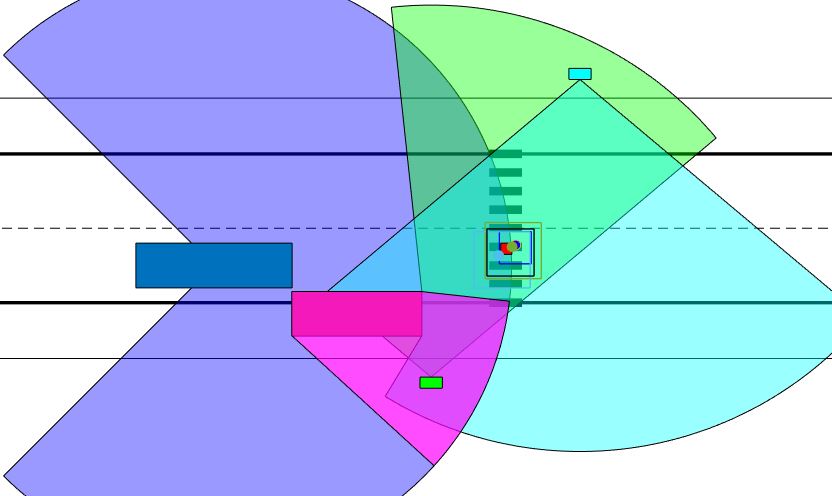}
        \caption{Scenario 3 with pedestrian.}
        \label{fig:scenario3a}
     \end{subfigure}
     \begin{subfigure}[b]{0.45\textwidth}
        \addtocounter{subfigure}{-1}
        \renewcommand\thesubfigure{\alph{subfigure}2}
        \centering
        \includegraphics[trim={1cm 1.5cm 0 0.5cm},clip,width=1\linewidth,height=0.7\linewidth]{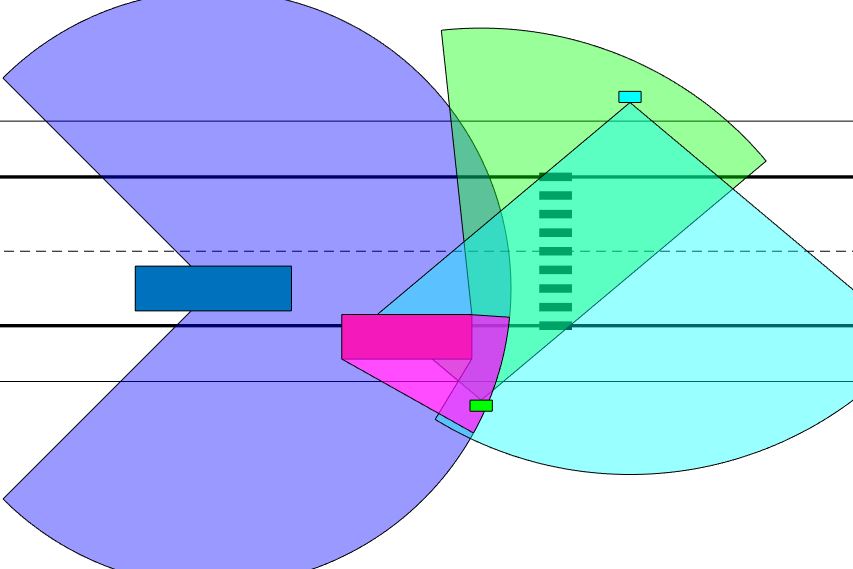}
        \caption{Scenario 3 without pedestrian.}
        \label{fig:scenario3b}
     \end{subfigure}
     \addtocounter{subfigure}{-1}
     \caption{Scenario 3 simulation results}
     \label{fig:scenario3}
    \end{subfigure}
    \caption{Simulation results for three scenarios.}
    \label{fig:scenariosresults}
\end{figure}

In Fig.~\ref{fig:scenariosresults}, the blue rectangle is the ego-vehicle with local senor field represented by the blue-shaded circle segment, and the occlusion is represented by the pink-shaded region. In these figures, local sensor measurements are represented by a blue dot, and unfilled blue squares represent respective zonotopic sets. The pedestrian is represented by a red square in  Figs.~\ref{fig:scenario1a},~\ref{fig:scenario2a} and \ref{fig:scenario3a}. In Figs.~\ref{fig:scenario1} and \ref{fig:scenario2}, yellow rectangle is the connected vehicle and in Figs.~\ref{fig:scenario2a} and \ref{fig:scenario2b} extended sensor is represented by the red-shaded circle segment. In Fig.~\ref{fig:scenario2a}, the red dot represents extended sensor measurement, and an unfilled red square represents its zonotopic set. Similarly, in Fig.~\ref{fig:scenario3}, extended sensors on road-side units 1 and 2 are represented by light green-shaded and light blue-shaded regions, respectively. Their measurements are represented by light green and light blue dots and zonotopic sets by unfilled squares. In Figs.~\ref{fig:scenario2a} and \ref{fig:scenario3a}, resultant zonotopes from the estimator are represented by unfilled black squares. 

\begin{figure*}[t]
    \centering
    \begin{subfigure}[b]{0.49\textwidth}
      \includegraphics[width=1\linewidth]{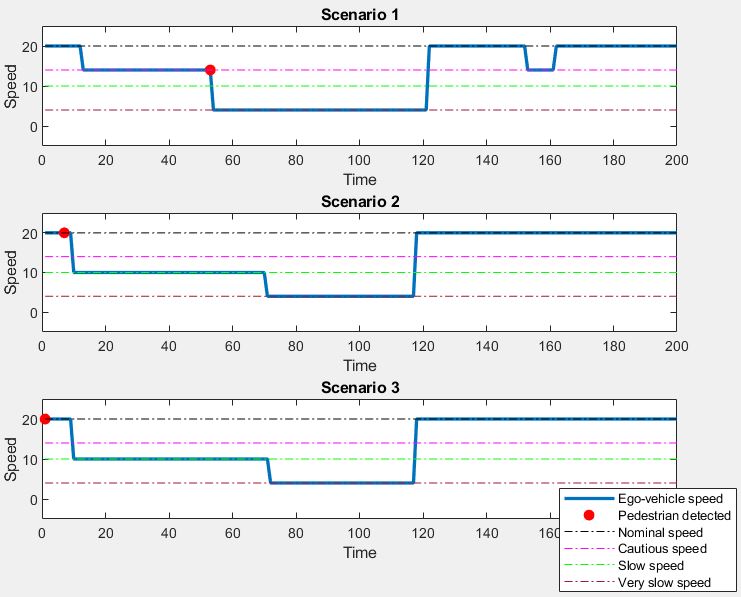}
    \caption{Three scenarios with pedestrian in the occluded region.}
    \label{fig:speed_with_Ped}
    \end{subfigure}
    \begin{subfigure}[b]{0.49\textwidth}
      \includegraphics[width=1\linewidth]{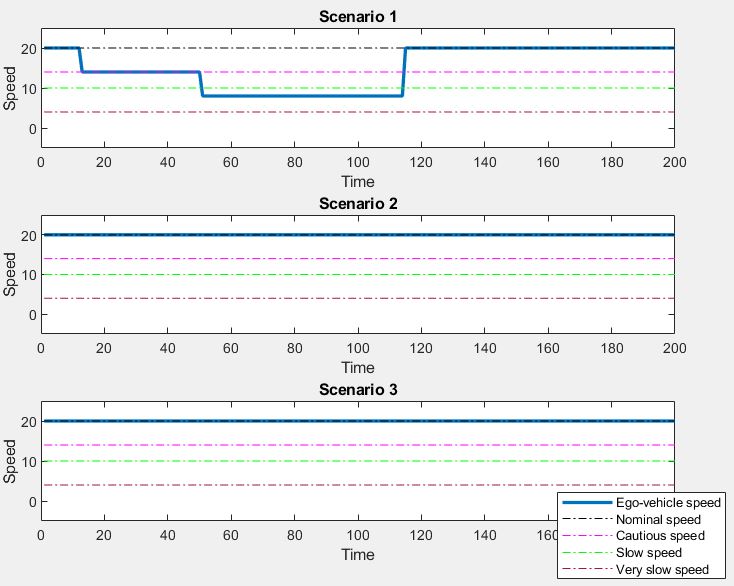}
    \caption{Three scenarios without pedestrian in the occluded region.}
    \label{fig:speed_without_Ped}
    \end{subfigure}
    \caption{Comparison of simulation results of ego-vehicle's speed for three scenarios which were stated in Section \ref{sec:pb}}
\end{figure*}

In Scenario~1, as there is an occlusion covering the side-walk, the ego-vehicle's speed is changed from nominal speed to cautious speed. In the case of having a pedestrian in the scenario, as shown in Fig.~\ref{fig:scenario1a}, the pedestrian was detected when it was too close to the ego-vehicle, so the speed was changed to very slow to avoid a collision. In the case where there was no pedestrian in the scenario as shown in Fig.~\ref{fig:scenario1b}, it is not certain whether there is a pedestrian until the ego-vehicle moves past the zebra crossing. Therefore as soon as the ego-vehicle detects zebra crossing, the speed is reduced to slow speed so that the ego-vehicle is vigilant. The recorded speed of the ego-vehicle in Scenario~1, for both the cases of with and without pedestrian, is plotted in Figs.~\ref{fig:speed_with_Ped} and \ref{fig:speed_without_Ped}, respectively. Furthermore, the average speed of the ego-vehicle in different scenarios is reported in Table.~\ref{tab:averagespeed_comparison}.

In Scenario~2, there is a connected vehicle moving in the opposite direction of the ego-vehicle. Moreover, this connected vehicle's extended sensor can cover the region that is occluded for the ego-vehicle. In the case of pedestrian, as shown in Fig.~\ref{fig:scenario2a}, the pedestrian has detected quick ahead of time than in Scenario~1 Fig.~\ref{fig:scenario1a}. Therefore, the speed is reduced to a slow level directly instead of a cautious level. In the case of Scenario~2 without pedestrian, as shown in Fig.~\ref{fig:scenario2b}, as the extended sensor well covers the occluded region, the ego-vehicle can accelerate with the nominal speed with certainty as the ego-vehicle is aware that there is no pedestrian in the occluded region. Therefore, share situational awareness has improved the performance, i.e., the average speed of the ego-vehicle as shown in Fig.~\ref{fig:speed_with_Ped} and \ref{fig:speed_without_Ped} and Table.~\ref{tab:averagespeed_comparison}.

It is not realistic to have a connected vehicle with an extended sensor covering the occlusion near the zebra crossing. For this reason, two extra road-side units have been installed, as shown in Scenario~3. In the case of a pedestrian in the scenario, as shown in Fig.~\ref{fig:scenario3a}, the pedestrian is detected a bit earlier than Scenario~2 as shown in Fig.~\ref{fig:scenario2a}. These road-side units are also advantageous when there is no pedestrian in the scenario, which can be seen in Table.~\ref{tab:averagespeed_comparison} where the average speed is equal to the nominal speed.

On comparing scenarios with and without shared situational awareness, i.e., Scenario~1 and Scenario~2~\&~3, improvement in the ego-vehicle performance is noticed. In a scenario with a pedestrian without the shared situational awareness, the pedestrian is detected close to the ego-vehicle. However, the pedestrian is detected in advance with shared situational awareness. In case there is no pedestrian, with shared situational awareness, the ego-vehicle's speed was maintained at the nominal speed level, which in turn improved the average speed of the ego-vehicle during the simulation. Moreover, with share situational awareness, the framework is supportive of detecting a sensor failure or false detection, which helps obtain accurate or reliable information regarding the surroundings.

\begin{figure}[h!]
    \centering
    \begin{subfigure}[b]{0.35\textwidth}
      \frame{\includegraphics[trim={0 0 1cm 0},clip,width=1\linewidth]{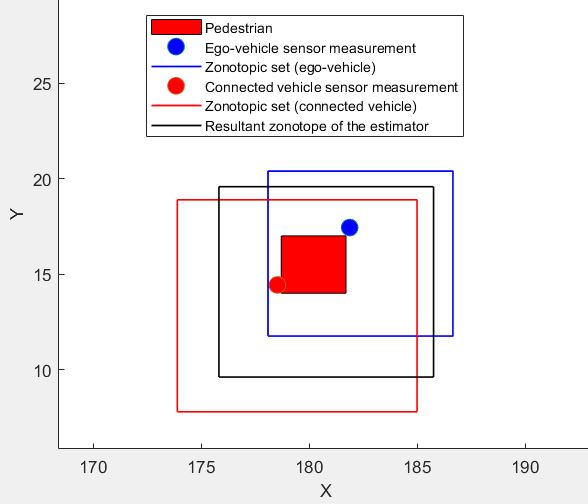}}
      \caption{Scenario 2 - estimated set zonotopes}
      \label{fig:scenario2_zono}
    \end{subfigure}
    \begin{subfigure}[b]{0.35\textwidth}
     \frame{\includegraphics[trim={0 0 1cm 0},clip,width=1\linewidth]{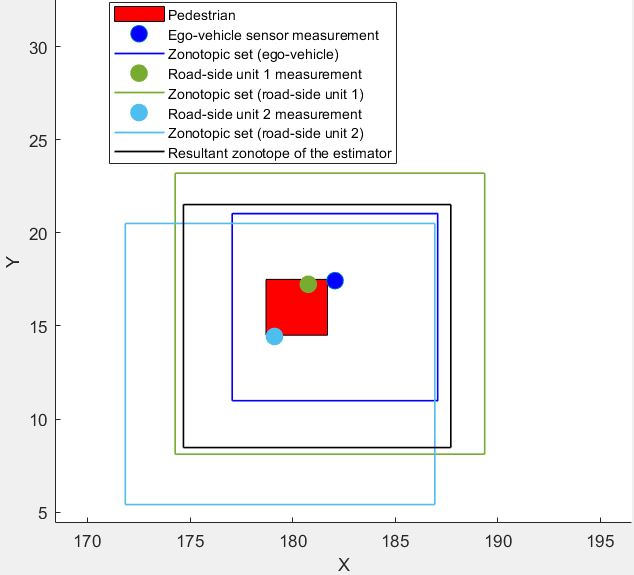}}
      \caption{Scenario 3 - estimated set zonotopes}
      \label{fig:scenario3_zono}
    \end{subfigure}
    \caption{Comparison of state estimation sets (zonotopes) for Scenario 2 and Scenario 3.}
    \label{fig:zonotope_comparision}
\end{figure}

In Fig.~\ref{fig:zonotope_comparision}, closer version of state estimation set zonotopes are presented for Scenario~2 shown in Fig.~\ref{fig:scenario2a} and Scenario~3 shown in Fig.~\ref{fig:scenario3a}. In Fig.~\ref{fig:scenario2_zono}, the volume of ego-vehicle's zonotopic set is $50.3$, the volume of connected vehicle's zonotopic set is $123.1$, and the volume of the resultant zonotope from the estimator is $94.0$. Similarly, in Fig.~\ref{fig:scenario3_zono}, the volume of ego-vehicle's zonotopic set is $100.5$ , the volume of road-side unit~1 zonotopic set is $227.5$, the volume of road-side unit~2 zonotopic set is $227.5$, and the volume of the resultant zonotope from the estimator is $170.0$. Therefore, the resultant zonotopes are an over-approximated zonotope, and it can also be observed that resultant zonotope is relatively smaller than the largest zonotopic set at a particular time step. 

\section{Conclusions}\label{sec:conc}

The proposed framework has the ability of sensor fusion of multiple sensors with varying uncertainties and estimates fusion from multiple vehicles or road infrastructures. It also provides guarantees along with the provided estimates, which are essential in safety-critical applications in general and autonomous vehicles in particular. The guarantees are based on set-membership estimation approaches. The potential of the proposed framework was illustrated with the aid of use cases along with simulation results. The framework also supports the detection of sensor failure, which is part of our future planned investigations. Real deployment on Scania autonomous vehicles is also part of future work. 


\section*{Acknowledgment}
This work was partially supported by the Wallenberg Artificial Intelligence, Autonomous Systems, and Software Program (WASP) funded by the Knut and Alice Wallenberg Foundation. It was also partially supported by the Swedish Research Council. 

\bibliographystyle{ieeetr}
\bibliography{ref.bib}
\end{document}